\documentclass[12pt]{article}
\usepackage{epsfig,amsmath}
\def\a{\alpha} 
\def\b{\beta}
          
\def\ga{\gamma}         
\def\gm{\Gamma}
\def\e{\eta} 
\def\la{\lambda}        

\def\m{\mu}
\def\n{\nu}

\def\eps{\epsilon}

\def\om{\omega}

\def\Dslash{{D\hspace{-8pt}/\hspace{3pt}}}
\def\pslash{{p\hspace{-5pt}/}}

\def\tp{\tilde{p}}

\def\idx{\int\! d^4\!x \,}

\def\ds{\displaystyle}

\begin{document}
\begin{titlepage}
\rightline{UCM-FT/00-13-02}

\vskip 1.5 true cm
\begin{center}
{\Large \bf Gauge-fixing independence of IR divergences in\\
non-commutative $\boldsymbol{U(1)}$, perturbative tachyonic\\
instabilities and supersymmetry\\}
\vskip 1.2 true cm 
{\rm F. Ruiz Ruiz}\footnote{E-mail: t63@aeneas.fis.ucm.es}\\ 
\vskip 0.3 true cm
{\it Departamento de F\'{\i}sica Te\'orica I, 
     Facultad de Ciencias F\'{\i}sicas}\\ 
{\it Universidad Complutense de Madrid, 28040 Madrid, Spain}\\
\vskip 1.2 true cm

{\leftskip=28pt \rightskip=38pt 
\noindent
It is argued that the quadratic and linear non-commutative IR
divergences that occur in $U(1)$ theory on non-commutative Minkowski
spacetime for small non-commutativity matrices $\theta^{\m\n}$ are
gauge-fixing independent. This implies in particular that the
perturbative tachyonic instability produced by the quadratic
divergences of this type in the vacuum polarization tensor is not a
gauge-fixing artifact. Supersymmetry can be introduced to remove from
the renormalized Green functions at one loop, not only the
non-logarithmic non-commutative IR divergences, but also all terms
proportional to $\theta^{\m\n}p_\n$.

\par
}
\end{center}

\vfil

\end{titlepage}
\setcounter{page}{2}


\noindent
In perturbative $U(1)$ gauge theory on non-commutative spacetime,
one-loop 1PI Green functions are known to have a non-analytic
behaviour for small values of the non-commutativity scale of
space-time \cite{Minwalla}.  The corresponding singularities are
called non-commutative IR divergences and occur in the non-planar
contributions of those Green functions whose planar parts are UV
divergent \cite{Minwalla}. This implies that one must expect quadratic
and logarithmic non-commutative IR divergences in the vacuum
polarization tensor, linear and logarithmic non-commutative IR
singularities in the gauge tree-vertex, and logarithmic
non-commutative IR divergences in the gauge four-vertex
\cite{Matusis}. In addition, and depending on the gauge choice, one
should also expect non-commutative IR divergences in the ghost Green
functions. The logarithmic non-commutative IR divergences in
particular satisfy a very important duality relation with the UV
divergences of the theory \cite{Minwalla}, in the sense that they can
be very simply read from one another (see below for an explicit
realization of this idea). By contrast, the quadratic and linear
non-commutative IR divergences have no counterpart as UV singularities
\cite{Matusis}. More importantly, their dependence on gauge-fixing is
not known, and thus one is not able to elucidate whether they are true
singularities or gauge-fixing artifacts. The purpose of this paper is
precisely to investigate their dependence on gauge-fixing. To do this,
we will consider the theory in an arbitrary Lorentz gauge and compute
the UV and non-commutative IR divergences at one loop. We will find
that only the two and three-point functions of the gauge field, but
not the ghost Green functions, contain quadratic and linear
non-commutative IR divergences, that these do not depend on the
gauge-fixing parameter and that they are equal to those previously
obtained in the Feynman background gauge \cite{Martin}. This suggests
that quadratic and linear IR divergences are independent of
gauge-fixing. Having argued this, we will then show that they lead to
a perturbative tachyonic instability for the theory. Finally, we will
explicitly illustrate how to cancel them with supersymmetry
\cite{Matusis}.

Non-commutative Minkowski spacetime is defined as the algebra
generated by position operators $X^\m$ satisfying commutation
relations $[X^{\m},X^{\n}] = i\theta^{\m\n}$, with $\theta^{\m\n}$ an
anti-symmetric real matrix. Indices are raised and lowered with the
Minkowski metric, for which we take $\eta_{\m\n}= {\rm
diag}(+,-,-,-)$. On this spacetime, a $U(1)$ gauge field is defined
\cite{Connes} as a real vector function $A^\m(x)$ with field strength
\begin{displaymath}
  F_{\m\n} = \partial_\m A_\n - \partial_\n A_\m 
           - i\,(A_\m\star A_\n - A_\n\star A_\m ) ~,
\end{displaymath}
where the symbol $\star$ denotes the Moyal product of functions, {\it
i.e.} 
\begin{displaymath}
  \big(f\star g\big) (x) = f(x)\, \exp\Big( \frac{i}{2}~
    \theta^{\m\n}\, \overleftarrow {\partial_\m} ~ 
      \overrightarrow{\partial_\n}\Big) \, g(x)~.
\end{displaymath}
The quantum $U(1)$ gauge theory can be defined through its Green
functions, whose generating functional is formally given in an
arbitrary Lorentz gauge by 
\begin{displaymath}
   Z[J,\bar{\eta},\eta] = \int [dA]\,[db]\,[d\bar{c}]\,[dc]~
     \exp\Big[\, iS + i\!\idx \big(\, J\,A + \bar{\eta}\,c 
                            + \bar{c}\,\eta \,\big)\, \Big] ~.
\end{displaymath}
Here $b$ is the auxiliary field imposing the Lorentz condition
$\partial\!A\!=\!f$, with $f$ an arbitrary function of $x^\m$,
$\bar{c}$ and $c$ are ghost fields, and
\begin{equation} 
   S = \idx \Big( -\,\frac{1}{4g^2}~ F^{\m\n}\!\star\! F_{\m\n}
          + \frac{\a}{2}~g^2\,b\!\star\! b - b\!\star\! \partial A
          + \bar{c}\!\star\!\partial_\m D^\m c \Big)
\label{class-action}
\end{equation}
is the gauge-fixed classical action. As usual, $g$ denotes the
coupling constant, $\a$ the gauge-fixing parameter, and $D_\m$ the
covariant derivative, whose action on a function $\om(x)$ is $D_\m\om
= \partial_\m \om - i \,(A_\m\star \om-\om\star A_\m ).$ In what
follows, to avoid running into inconsistencies with unitarity
\cite{Gomis}, we restrict ourselves to matrices $\theta^{\m\n}$ such
that $\theta^{0i}=0$ for $i=1,2,3$. This choice is called magnetic and
ensures that the field theory exists as the zero slope limit of a
string theory \cite{Seiberg}.

At one loop, the planar parts of only the vacuum polarization tensor
$i\Pi_{\m\n}(p)$, the gauge three-vertex
$i\gm_{\m_1\m_2\m_3}(p_1,p_2,p_3)$, the gauge four-vertex
$i\gm_{\m_1\m_2\m_3\m_4} (p_1,p_2,p_3,p_4)$, the ghost self-energy
$\Pi^{\bar{c}c}(p)$ and the ghost vertex $i\gm^{\bar{c}Ac}_\m
(p_1,p_2,p_3)$ are UV divergent. Therefore one should expect
non-commutative IR singularities in only these Green functions. To
regularize the theory, we use dimensional regularization with
$D=4+2\eps$. When sending $\eps\to 0$, two types of terms appear:
$1/\eps$ pole terms and finite terms.  The pole terms, being
divergent, we keep them. The finite terms, as functions of
$\theta^{\m\n}$, develop singularities for $\theta^{\m\n}\to 0$. These
singularities are our main interest here. We have computed the double
limit ${\ds \lim_{\theta\to 0}\,\lim_{\eps\to 0}}$ of these five Green
functions and obtained, modulo finite contributions:
\begin{eqnarray}
  &&i\Pi_{\m\n}(p)  
     = \frac{2i}{\pi^2} ~ \frac{\tp_\m\tp_\n}{(p\circ p)^2}    
     - \frac{i}{16\pi^2}  \Big(\frac{13}{3}-\a\Big)
       \Big[\,\frac{1}{\eps} - \ln\,(p^2\,p\!\circ\! p) \,\Big]\,
       (p^2\eta_{\m\n} - p_\m p_\n)  \label{2-point}\\[12pt]
  && {\ds i\gm_{\m_1\m_2\m_3}(p_1,p_2,p_3)
      = \frac{2}{\pi^2}\, \cos\left(\frac{p_1\wedge p_2}{2}\right)
         \sum_{i=1}^3\> \frac{(\tp_i)_{\m_1} (\tp_i)_{\m_2} (\tp_i)_{\m_3}}  
                           {(p_i\circ p_i)^2} } \nonumber \\
  && \hphantom{i\Pi_{\m\n}(p)} 
  {\ds +\, \frac{1}{16\pi^2}\, 
       \sin\!\left(\frac{p_1\wedge p_2}{2}\right) 
        \Big(\frac{17}{3}- 3\a\Big)\,
        \Big[ \frac{1}{\eps} - \frac{1}{3} \>\sum_{i=1}^3\, 
             \ln\,(p_i\circ p_i) \,\Big]  } \nonumber \\
  && \hphantom{i\Pi_{\m\n}(p)} 
  {\ds \times\, \Big[ \eta_{\m_1\m_2}(p_1-p_2)_{\m_3}
           + \eta_{\m_2\m_3}(p_2-p_3)_{\m_1}
           + \eta_{\m_3\m_1}(p_3-p_1)_{\m_2}\Big] } 
                                 \label{3-point} 
\end{eqnarray}
\begin{eqnarray}
  && {\ds i\gm_{\m_1\m_2\m_3\m_4}(p_1,p_2,p_3,p_4) 
         = \frac{i}{16\pi^2} } \nonumber \\ 
  && \hphantom{i\Pi_{\m\n}(p)}
  {\ds \times \bigg[ 4\,
       \big( \e_{\m_1\m_3}\e_{\m_2\m_4}-\e_{\m_1\m_4}\e_{\m_2\m_3} \big)
              \sin\!\left( \frac{p_1\wedge p_2}{2}\right)
              \sin\!\left( \frac{p_3\wedge p_4}{2}\right) 
              f_\eps(\tp_1,\tp_2) } \nonumber \\
  && \hphantom{i\Pi_{\m\n}(p)~}
  {\ds +\,4\, 
       \big( \e_{\m_1\m_2}\e_{\m_3\m_4}-\e_{\m_1\m_4}\e_{\m_2\m_3} \big)
           \sin\!\left( \frac{p_1\wedge p_3}{2}\right)
           \sin\!\left( \frac{p_2\wedge p_4}{2}\right) 
              f_\eps(\tp_1,\tp_3) } \nonumber \\
  && \hphantom{i\Pi_{\m\n}(p)~}  
  {\ds +\, 4\,
       \big( \e_{\m_1\m_2}\e_{\m_3\m_4}-\e_{\m_1\m_3}\e_{\m_2\m_4} \big)
           \sin\!\left( \frac{p_1\wedge p_4}{2}\right)
           \sin\!\left( \frac{p_2\wedge p_3}{2}\right) 
              f_\eps(\tp_1,\tp_4) \bigg] } \label{4-point} \\[12pt]
  && i\Pi^{\bar{c}c}(p) = \frac{i}{16\pi^2}\> \frac{3-\a}{2}\>
     \Big[\, \frac{1}{\eps} - \ln\,(p\circ p)\,\Big]\>p^2 
                            \label{ghost-self}\\[12pt]
  && i\gm^{\bar{c}Ac}_\m(p_1,p_2,p_3) = -\>\frac{\a}{16\pi^2}\>
     \sin\!\left(\frac{p_2\wedge p_3}{2}\right)    \nonumber\\
  && \hphantom{i\Pi_{\m\n}(p)}
     \times \Big[\, \frac{1}{\eps} + \ln(p_1\circ p_1) 
               - \ln(p_2\circ p_2) -  \ln(p_3\circ p_3) \Big]\>(p_1)_\m 
                             \label{ghostvertex} 
\end{eqnarray}
Here $\tp^\m$, $p_i\wedge p_j$ and $p\circ p$ denote 
\begin{displaymath}
  \tp^\m = \theta^{\m\n}p_\n \qquad 
  p\wedge q = \theta^{\m\n} p_\m q_\n \qquad 
  p\circ p = -\, \theta^{\m\n}\,\theta_{\m}^{~\,\tau}\,p_\n\, p_\tau ~,
\end{displaymath}
and the function $ f_\eps(\tp_j,\tp_k)$ is given by 
\begin{displaymath}
\begin{array}{l}
  {\ds  f_\eps(\tp_j,\tp_k) =
           \frac{1}{\eps}\>\Big(\frac{4}{3}-2\a\Big)  
        +  \frac{1}{8}\, \Big[\, (\a+3)\,(\a-1) + \frac{31}{3}\,\Big]\>
           \sum_{i=1}^4\, \ln(p_i\circ p_i) }\\[12pt]
\hphantom{ f_\eps(\tp_j,\tp_k)\>}
  {\ds -\> \frac{1}{2}\>\big[ 9+(1-\a)^2\big]\>
       \ln\big[(p_j+p_k)\circ(p_j+p_k)\big] ~.}
\end{array}
\end{displaymath}
In the ghost vertex $i\gm^{\bar{c}Ac}_\m(p_1,p_2,p_3)$, $p_1^\m,
~p_2^\m$ and $p_3^\m$ are the momenta of the incoming ghost, the gauge
field and the outgoing ghost, respectively. The $1/\eps$ pole terms in
these equations were first calculated in ref. \cite{beta1}, give for
the one-loop beta function $\b_1=-11/3$ \cite{beta1} \cite{beta2} and
are subtracted by adding UV counterterms. This UV renormalization
procedure leaves the non-commutative IR divergences unchanged, a
feature observed in all field theories considered so far
\cite{several}. 

We observe that the first term in the vacuum polarization tensor
(\ref{2-point}) diverges quadratically for $\theta^{\m\n}\!\to 0$ and
that the first term in the gauge three-vertex (\ref{3-point}) does it
linearly. More importantly, their coefficients do not depend on the
gauge-fixing parameter $\a$ and are equal to those obtained in the
Feynman background gauge \cite{Martin}. We would like to emphasize
that this $\a$-independence is not trivial. To illustrate this,
suffice to say that the Feynman diagrams that contribute to {\it e.g.}
the gauge three-vertex give linear non-commutative IR divergent
contributions of type $\,\cos(\frac{1}{2}\,p_1\wedge p_2)~ \eta_{\m_1
\m_2}\> \big[ (\tp_i)_{\m_3}/(p_i\circ p_i)\big]\,$ and
$\,\cos(\frac{1}{2}\,p_1\wedge p_2)\> \big[ (\tp_i)_{\m_1}
(\tp_i)_{\m_2} (\tp_i)_{\m_3}/ (p_i\circ p_i)^2\big]$ whose
coefficients depend on $\a$, and that only after summation over
diagrams these contributions cancel except for the first term in
eq. (\ref{3-point}). We also note that the ghost self-energy does not
have linear non-commutative IR divergences, even though on general
grounds it is susceptible to them. All this suggests that the
quadratic and linear non-commutative IR divergences in the theory at
one loop are gauge-fixing independent. By contrast, the logarithmic
terms $\ln(p\circ p)$ depend on $\a$ and are in complete accordance
with UV/IR duality. Indeed, in the non-commutative IR momenta domain
$|\tp_i|\sim\theta\Lambda_{\rm IR}\to 0$, the terms in a Green
function with logarithmic dependence on $\theta^{\m\n}$ can be
obtained from the pole terms in the same Green function through the
identifications $1/\eps \leftrightarrow \ln\Lambda_{\rm UV}^2$,
$\Lambda_{\rm UV}\leftrightarrow 1/\theta\Lambda_{\rm IR}$.

Now that we have argued that the quadratic non-commutative IR
divergences are independent of the choice of gauge, we discuss their
relevance for perturbative tachyonic instability. We use the notation
$p^\m=(E,\vec{p})$ and consider momenta in the domain $\,g^2/(p\circ
p)\,|\vec{p}|^2 \ll 1$, where perturbation theory is valid. In this
domain, the leading behaviour of the renormalized vacuum polarization
tensor in any admissible renormalization scheme is given by
\begin{displaymath}
  i\Pi^{\rm ren}_{\m\n}(p) = -\,\frac{i}{g^2}~ 
    (p^2\eta_{\m\n} - p_\m p_\n) 
   + \frac{ia}{\pi^2} ~ \frac{\tp_\m\tp_\n}{(p\circ p)^2}\, 
   +\, {\rm subleading~contributions}~,
\end{displaymath}
where, according to eq. (\ref{2-point}), $a=2$. For a ``photon'' of
momentum $p^\m$ polarized in the direction of $\tp^\m$, with plane
wave field $f(p)\,\tp^\m e^{ipx}$, the dispersion relation takes the
form
\begin{equation}
  E^2 = |\vec{p}|^2 - \frac{a}{\pi^2}~\frac{g^2}{p\circ p}~.
\label{dispersion}
\end{equation}
Since $a=2$ and $p\circ p$ is positive, this gives rise to a
perturbative tachyonic instability. Note that the existence of this
instability depends on the sign of $a$; had $a$ been negative, there
would not be such instability. Instabilities of this type have also
been found in scalar theories on non-commutative space-times
\cite{Landsteiner}. The quadratic and linear non-commutative IR
divergences under consideration are also a source of problems for UV
renormalizability at higher loops, since a one-loop non-commutative IR
divergence nested in a two-loop diagram produces IR non-integrable
singularities which make renormalization at two loops unlikely. All in
all, one would like to find a way to cancel these non-commutative IR
divergences. It has been argued in ref. \cite{Matusis} that in the
Feynman Lorentz Gauge this can be done by introducing
supersymmetry. The gauge-fixing independence of the divergences under
consideration, together with the observation that the coupling of
fermions to the gauge field is through the covariant derivative and
thus does not depend on gauge-fixing, makes the cancellation picture
consistent. Let us explicitly realize this cancellation mechanism for
$N=1$ supersymmetry.

The coupling of a Majorana spinor $\la$ to the gauge field $A^\m$ adds
a term $\,-\frac{1}{2g^2}\,\bar{\la} \star \Dslash \la$ to the
classical action. Since the field $\la$ must belong to the same
supermultiplet as the gauge field, it must transform according to the
adjoint representation and thus for the covariant derivative we must
take $D_\m\la = \partial_\m \la - i \,(A_\m\star \la-\la\star A_\m
)$. At one loop, this coupling gives additional contributions to the
two, three and four-point 1PI functions of the gauge field. For
brevity, we only display here the first two:
\begin{eqnarray}
  &&\Delta\big[ i\Pi_{\m\n}(p)\big]  
     = -\frac{2i}{\pi^2} ~ \frac{\tp_\m\tp_\n}{(p \circ p)^2}    
     - \frac{i}{16\pi^2} ~\frac{4}{3}
       \Big[ -\,\frac{1}{\eps} + \ln\,(p^2 p\!\circ\! p) \,\Big]\,
       (p^2\eta_{\m\n} - p_\m p_\n) \label{2-Maj}\\[12pt]
  && {\ds \Delta\big[ i\gm_{\m_1\m_2\m_3}(p_1,p_2,p_3)\big]
      = - \frac{2}{\pi^2}\, \cos\left(\frac{p_1\wedge p_2}{2}\right)
         \sum_{i=1}^3\> \frac{(\tp_i)_{\m_1} (\tp_i)_{\m_2} (\tp_i)_{\m_3}}  
                           {(p_i\circ p_i)^2} } \nonumber \\
  && \hphantom{i\Pi_{\m\n}(p)} 
  {\ds +\, \frac{1}{16\pi^2}\, 
       \sin\!\left(\frac{p_1\wedge p_2}{2}\right) 
        \frac{8}{3}\,
        \Big[-\,\frac{1}{\eps} + \frac{1}{3} \>\sum_{i=1}^3\, 
             \ln\,(p_i\circ p_i) \,\Big]  } \nonumber \\
  && \hphantom{i\Pi_{\m\n}(p)} 
  {\ds \times\, \Big[ \eta_{\m_1\m_2}(p_1-p_2)_{\m_3}
           + \eta_{\m_2\m_3}(p_2-p_3)_{\m_1}
           + \eta_{\m_3\m_1}(p_3-p_1)_{\m_2}\Big] ~.} 
                                 \label{3-Maj} 
\end{eqnarray}
We see that indeed these two contributions cancel the quadratic and
linear non-commutative IR divergences in eqs. (\ref{2-point}) and
(\ref{3-point}). Although we are not displaying partial results here,
we mention that the entire contribution to the vacuum polarization
tensor proportional to $\tp_\m\tp_\n$, and not only its divergent part
for $\theta^{\m\n}\to 0$, is canceled by the contribution proportional
to $\tp_\m\tp_\n$ from the Feynman diagram with Majorana fermions
flowing along the loop. Hence the vacuum polarization tensor for the
supersymmetric theory is, for arbitrary values of $\theta^{\m\n}$,
proportional to $p^2\eta_{\m\n}-p_\m p_\n$. This implies that there is
no momentum region for which a tachyonic instability could arise at
one loop. Analogously, all terms in the gauge three-vertex
proportional to any $\tp_i^\m$ $(i=1,2,3)$, and not only their
divergent parts for $\theta^{\m\n}\to 0$, are canceled by
supersymmetry. All this does not guarantee, however, that the
supersymmetric theory is free of non-logarithmic non-commutative IR
divergences at one loop, since linear non-commutative IR singularities
may still occur in the Majorana fermion self-energy $\Sigma(p)$. To
check if this is so we have calculated $\Sigma(p)$ and obtained that
it has no contribution at all proportional to $\ga^\m\tp_\m$ and that
in the double limit of interest it reduces to
\begin{displaymath}
  \Sigma(p) =  - \frac{\a}{16\pi^2} \Big[\,\frac{1}{\eps} 
                     - \ln\,(p^2\,p\!\circ\! p) \,\Big]\,\pslash~,
\end{displaymath}
which only contains logarithmic non-commutative IR divergences. In
summary, in the supersymmetric theory all non-commutative IR
divergences are logarithmic and the tensor structures that enter the
one-loop 1PI Green functions are the same as for the conventional
commutative case, thus increasing the expectations for
renormalizability at higher loops. Note finally that supersymmetry
modifies the beta function, which now is $\b_1= -3$.

We conclude summarizing our results. We have argued that the quadratic
and linear non-commutative IR divergences in $U(1)$ gauge theory on
non-commutative spacetime are gauge-fixing independent. From the point
of view of local perturbative quantum field theory, the presence of
non-logarithmic non-commutative IR divergences is problematic.
Indeed, a renormalized one-loop 1PI Green function with divergences of
this type, when nested in a two-loop diagram, gives rise for small
$\theta^{\m\n}$ to non-integrable singularities, thus making unlikely
UV renormalization at higher loops. In addition, the quadratic
non-commutative IR divergences in the vacuum polarization tensor yield,
for polarizations parallel to the non-commutative directions, a
tachyonic dispersion relation which introduces a perturbative
tachyonic instability. One way to remove these non-commutative IR
divergences and the problems they bring along is is to introduce
supersymmetry, since 1PI Feynman diagrams with Majorana fermions
flowing along the loop provide the necessary contributions to cancel
the quadratic and linear non-commutative IR divergences in the Green
functions of the gauge field. Further, since the self-energy of the
Majorana fermion does not contain divergences of this type, the
supersymmetric theory is free of quadratic and linear non-commutative
IR divergences. Thus in a superfield approach, one expects to only
find UV and logarithmic non-commutative IR divergences \cite{Zanon}.

\section*{Acknowledgment}

The author is grateful C.P. Mart\'{\i}n for many conversations and
discussions, and to CICyT, Spain, for financial support through grant
No. PB98-0842.

\end{document}